\documentclass[12pt,preprint]{aastex}
\usepackage{emulateapj5}
\shorttitle{$\Omega_M=0.33\pm 0.035$}
\shortauthors{Turner}

\begin{document}
\title{The Case for $\Omega_M = 0.33\pm 0.035$}

\author{Michael S. Turner}
\affil{Departments of Astronomy \& Astrophysics and of Physics, Enrico Fermi Institute\\
and Center for Cosmological Physics, The University of Chicago, Chicago, IL~~60637-1433}
\and
\affil{NASA/Fermilab Astrophysics Center\\
Fermi National Accelerator Laboratory, Batavia, IL~~60510-0500}
\email{mturner@oddjob.uchicago.edu}

\begin{abstract}
For decades, the determination of the mean density of matter
($\Omega_M$) has been tied to the distribution of light.
This has led to a ``bias,'' perhaps as large as a factor of
2, in determining a key cosmological parameter.
Recent measurements of the physical properties of clusters,
cosmic microwave background (CMB) anisotropy and the power spectrum of
mass inhomogeneity now allow a determination of $\Omega_M$
without ``visual bias.''  The early data
lead to a consistent picture of the matter and baryon densities,
with $\Omega_B = 0.039\pm 0.0075$ and $\Omega_M =
0.33\pm 0.035$.
\end{abstract}


\section{Introduction}

The mean mass density of the Universe is a cosmological
parameter of great importance.  It can be expressed as a fraction of
the critical density, $\Omega_M \equiv \rho_M /\rho_{\rm crit}$,
or in physical units (${\rm g\,cm^{-3}}$).
Moving between the Hubble constant:
\begin{equation}
\rho_M = 1.88(\Omega_M h^2)\times 10^{-29}\,{\rm g\ cm^{-3}}
\end{equation}
where as usual, $H_0=100h\,{\rm km\,sec^{-1}\,Mpc^{-1}}$
and $\rho_{\rm crit} = 3H_0^2/8\pi G = 1.88h^2\times 10^{-29}\,
{\rm g\ cm^{-3}} \approx 10^{-29}\,{\rm g\,cm^{-3}}$.

Over the past thirty years much effort has been devoted to
determining the mean matter density  (see e.g., Faber \& Gallagher, 1979;
Dekel et al, 1997; Turner, 2000; Primack, 2001).  The task is daunting:
the density of matter in a large enough sample of the Universe to
be representative ($> 10^6 \,{\rm Mpc}^3$) must be determined.  The most well
developed techniques have been tied in one way or another
to the distribution of light; e.g., mass-to-light ratios and peculiar 
velocities (see e.g., Bahcall et al,
1995; Carlberg et al, 1997; Dekel, 1994; Willick \& Strauss, 1998).

The mass-to-ratio technique begins from the deceptively simple equation that
relates the mean luminosity density and the mean mass density:
\begin{equation}
\rho_M = \langle M/L \rangle {\cal L}.
\end{equation}
Using the canonical value for the B-band luminosity density \
today\footnote{Important questions still remain about the mean luminosity density,
including evolution.} and solar units, Eq. 2 can be written as the familiar,
\begin{equation}
\Omega_M = \langle M/L \rangle_B /1200h .
\end{equation}

The task is now transformed to determining the mean mass-to-light
ratio.  Either $\langle M/L \rangle$ must be measured in a large enough
volume to be representative ($>10^6\,{\rm Mpc}^3$),
or measured for a system that can be argued to be ``typical.''
Needless to say, the accuracy of the inferred mean
matter density can be no better than that of the mean mass-to-light ratio.

In attempting to determine $\langle M/L \rangle$ attention has been
focused on clusters, because of their size and relatively well determined
total masses.  However, clusters are not large enough to provide
a representative sample of the Universe, and because the cluster
environment is a high density one
($10^2-10^3$ times the mean density) in which only a few percent of galaxies
find themselves, there is no {\it a priori} reason to believe that their
mass-to-light ratio is representative.

The CNOC sample provides the largest and best understood
sample of clusters, and from it Carlberg et al (1997) infer
$\Omega_M = 0.19\pm 0.04$.  By comparing cluster data
with N-body simulations Bahcall et al (2000) find a similar value,
$\Omega_M = 0.16\pm 0.05$.

Field galaxies are more representative, but
problematic because galaxy halos are large.  In fact,
no definitive evidence yet exists that the total mass
of even a single galaxy halo has been determined.
The advent of weak-gravitational lensing to probe the far reaches
of galaxy halos and to determine their total masses may soon
surmount this hurdle (see e.g., Fischer et al, 2000).  However, for now,
the values for $\Omega_M$ inferred from clusters (quoted above)
represent the state of this art in mass-to-light ratios.

Peculiar velocities and bulk flows are directly tied to the underlying
mass distribution; comparison of peculiar velocities and
mass inhomogeneity ($\delta \rho/\rho $) fixes
the mean density (see e.g., Dekel 1994).  However, it is the
distribution of galaxies -- not mass -- that is measured
in redshift surveys.  For this reason, this technique
probes $\beta = \Omega_M^{0.6}/b$, where $b$ is the
assumed linear bias factor between light and mass.  (Nonlinear
effects break the degeneracy, but there has been little
success in exploiting this fact.)  The peculiar-velocity technique
has its own convergence problem:  Unless the mass distribution
is surveyed to a volume sufficiently large to include all of
the inhomogeneity that gives rise to the peculiar velocity,
an accurate value for $\Omega_M$ will not be obtained.  Currently,
there are unexplained differences in the values obtained
for $\beta$, which range from 0.4 to almost 1.

A first step toward a physically based technique was proposed by White et al (1993).
The method goes like this.  Assume that galaxy clusters provide a ``fair sample'' of
matter in the Universe, i.e., that their composition reflects the
universal baryon-to-total mass ratio
($= \Omega_B/\Omega_M$).  Then, from our knowledge of the
baryon density from big-bang nucleosynthesis (BBN), the mean matter density
can be inferred:
\begin{equation}
\Omega_M = \left( {M_T\over M_B} \right)_{\rm cluster} \Omega_B({\rm BBN})
\end{equation}
While the bias factor $b$ never enters, there are potential
systematics:  Can one find all the cluster baryons?  Can the ratio
be determined on large enough scales to ensure that baryon
settling is not an issue?  White et al (1993) and others
(e.g., Mohr et al, 1998) have argued that the answer to these
questions is yes, and several recent recent reviews of
the mean matter density have given significant weight to this technique
(see e.g., Dekel et al, 1997; Turner, 2000; and Primack, 2001).

Today, other physical measurements, including CMB anisotropy and
measurements of the power spectrum of matter inhomogeneity, can be
added to the mix.  As I will describe, the physically based methods
have become mature enough to self-consistently and robustly
pin down the mass density with a relatively small error bar.
The value I derive from the current data is:
\begin{equation}
\Omega_M = 0.33 \pm 0.035\ \ \ (1\ \sigma )
\end{equation}
There are however several potential sources of systematic error
(see below).  Nonetheless, I believer that this marks an important
first step toward an accurate determination of the mean matter
density that is independent of the relationship of mass to light.

\section{The Input Data:  Physical Measurements}

A suite of physical measurements can now probe the distribution of matter and
the mean matter in an unbiased way.  They include accurate determinations of
CMB anisotropy on angular scales down to 0.1 degrees and measurements
of the shape of the power spectrum of matter inhomogeneity.
CMB anisotropy depends significantly
upon the physical baryon and matter density, as well as other parameters,
but not upon the connection between mass and light.

The shape of the power spectrum of mass inhomogeneity
depends significantly upon $\Omega_Mh$ and $\Omega_M/\Omega_B$;
it can be measured by weak-gravitational lensing (cosmic shear)
and through the distribution of galaxies (redshift surveys).  Cosmic shear
measurements are insensitive to bias, but are much less mature.
Redshift surveys are sensitive to bias in determining
the power spectrum itself.
However, the shape of the power spectrum is only sensitive to strong
scale-dependent bias on large scales around the bend in the power spectrum
associated with matter-radiation equality ($\sim 30\,{\rm Mpc}/\Omega_Mh^2$).

With MAP, Planck, full results from the SDSS and 2dF, and additional
cosmic-shear measurements coming, the future for an accurate,
physically based measurement of $\Omega_M$
is very bright.  I believe there is now enough data to make a preliminary
estimate of $\Omega_M$ using these techniques.

The following physical measurements comprise my input data for determining
the baryon and total matter densities:
\begin{eqnarray}
{\rm Power\ Spectrum}\ \ \ \ \  \Omega_M h & = &  0.20\pm 0.03        \nonumber \\
\Omega_B/\Omega_M & = & 0.15\pm 0.07                               \nonumber \\
{\rm CMB\ Anisotropy}\ \ \  \Omega_Mh^2 & = & 0.16 \pm 0.04   \nonumber \\
  \Omega_Bh^2 & = & 0.022^{+0.004}_{-0.003}                        \nonumber \\
{\rm BBN}\qquad\qquad\qquad\ \, \Omega_Bh^2 & = & 0.020 \pm 0.001     \nonumber \\
{\rm Clusters}\qquad\qquad      \Omega_B/\Omega_M & = & (0.07 \pm 0.007)h^{-3/2}
\ \ {\rm (X\ ray)}                                                 \nonumber \\
            & = & (0.08 \pm 0.01)h^{-1} \ \ {\rm (S-Z)}          \nonumber \\
{\rm Hubble\ constant}\qquad \ \ \ \ h & = & 0.72 \pm 0.07                \nonumber
\end{eqnarray}

The shape of the power spectrum of matter inhomogeneity depends
upon $\Omega_M h$ and $\Omega_B/\Omega_M$, as well as other parameters.
Currently, the best determination of these two parameters
come from an analysis of 160,000 redshifts in
the 2-degree Field Galaxy Redshift Survey (2dFGRS; Percival et al, 2001).
The baryon-to-total matter ratio is determined from the presence of
``baryon bumps'' in the power spectrum; it is hardly a significant
result at the moment and gives almost no weight to the determination
of $\Omega_M$ and $\Omega_B$.
Early results from the Sloan Digital Sky Survey (SDSS) are consistent
with the 2dFGRS numbers, $\Omega_M h = 0.19 \pm 0.04$ (Dodelson et al., 2002;
Szalay et al., 2001).

Significant improvement in both $\Omega_M h$ and $\Omega_B/\Omega_M$
can be expected when the 2dF and SDSS surveys have
amassed and analyzed their full samples of 250,000 and 600,000
redshifts respectively.  The results will also be more robust.
For example, current analyses assume scale-invariant
perturbations ($n=1$) and there are still correlations between
$\sigma_8$ and $\Omega_Mh$.  While cosmic-shear measurements of
large-scale structure are not yet good enough
to have significant leverage on $\Omega_M$ and $\Omega_B$, they
will be in the future (see e.g., van Waerbeke et al.,
2001; Maoli et al., 2001).

The structure of the acoustic peaks in the CMB angular power spectrum
depends upon the total matter density and the baryon density
(in physical units):  the ratio
of the odd to even peaks pins down the baryon density and the height of
the first peak is sensitive to the total matter density (see e.g.,
Hu et al, 1997).  I have used the values extracted from the first-year
DASI results (Pryke et al, 2002).  (The correlations between the two densities are small.)
A re-analysis of the 1998 flight of Boomerang yields
similar values (Netterfield et al, 2002).  Combining the two results --
which I have not done -- would reduce the error bars by about $\sqrt{2}$.

While the consistency of DASI, a radio
interferometer operating at around 30 GHz, and Boomerang,
bolometers in bands above 100 GHz, is very reassuring,
there are still significant
issues.  Both groups used a limited number of parameters (7)
and priors on the Hubble constant ($h> 0.45$) and optical
depth to last scattering ($\tau < 0.4$).  The parameters preclude
a large contribution by tensor (gravitational
wave) perturbations; while unlikely, relaxing that assumption
can significantly change the inferred baryon density (see e.g.,
Wang, Tegmark and Zaldarriaga, 2002).  With MAP and Planck
on the horizon, significant improvement, both in the precision
and the robustness of the results, lies ahead.

At present, the most precise determination of the physical baryon density comes
from combining measurements of the primeval abundance of deuterium (see e.g.,
O'Meara et al, 2001) with accurate theoretical predictions of the light-element
abundances (see e.g., Burles et al, 2001).
I follow the analysis of Burles et al (2001) in adopting $\Omega_Bh^2 =
0.02\pm 0.001$.  Here too, there are still issues to be resolved:  possible
unidentified systematics in the determination of the primeval deuterium
abundance; reliable values for the primeval $^4$He and $^7$Li abundances
to compare with the values predicted from the deuterium-determined
baryon density.  Over the next decade, as more deuterium systems
are found and the $^4$He and $^7$Li abundances are better understood,
the accuracy and reliability of the BBN baryon density should improve.

The baryon-to-matter ratio in clusters can be determined by x-ray measurements
alone and by measurements of the Sunyaev -- Zel'dovich (SZ) distortion of the CMB
combined with x-ray measurements.  For x-ray measurements alone, I adopt the
cluster baryon fraction determined from a sample of 45 clusters
by Mohr et al (1998), and for the SZ/X-ray determination I use the
cluster sample of Grego et al (2001).

Potential sources of systematic error remain.  There is
some clumping of cluster gas, which could lead to an overestimation
of the amount of gas.  The baryon-to-total mass ratio does vary
with radius, and the calibrations are done by comparison with
numerical simulations.  The treatment of gas dynamics in these
simulations still involves approximations and assumptions.  Better
simulations, more observations of clusters with the SZ technique
and x-ray should address the key issues and improve the reliability
of this sampling technique.

The sensitivity of the abundance of clusters, both as a function of cluster mass
and redshift, to the matter density has been used to estimate $\Omega_M$
(see e.g., Bahcall \& Fan, 1998; Blanchard et al, 2000; Henry, 2000).
However, the range in inferred values is broad, $\Omega_M =0.1 - 1$,
largely because of the exponentially important, but
uncertain relation between cluster mass and x-ray temperature.  For this
reason, I do not include these measurements in my
determination of the matter density.

Finally, the value of the Hubble constant is important in
converting physical densities to fractions of critical density;
I adopt the value determined by the Hubble Key Project:  $h=0.72\pm 0.07$
(Freedman et al, 2001).  The error is almost entirely due to systematics
(the statistical error is only $\sigma_h = \pm 0.02$).  The
largest part of the systematic error budget is in the distance
to the Large Magellanic Cloud.  Possible dependence of the Cepheid
period -- luminosity relation on metallicity may also be an issue.

\section{The case for $\Omega_M = 0.33 \pm 0.035$}

Having adopted these physical measurements as input data,
it is straightforward to deduce the fractions
of critical density contributed by matter and by baryons,
as well as the Hubble constant.  But first, consider
the consistency of these measurements.  The physical baryon density
($\Omega_Bh^2$) is determined by BBN, CMB anisotropy and the power
spectrum + Hubble constant:
\begin{eqnarray}
\Omega_B h^2   & = & 0.020 \pm 0.001\qquad {\rm BBN}\nonumber \\
                & = & 0.022^{+0.004}_{-0.003} \qquad\ \ \ {\rm CMB}\nonumber\\
                & = & 0.022\pm 0.011 \qquad {\rm Power\ Spectrum\ }+\ H_0
\end{eqnarray}
These three, independent determinations of the baryon density are clearly consistent,
giving one confidence in the case for  a low baryon density
($\rho_B \approx 4\times 10^{-31} \,{\rm g\ cm^{-3}}$).  They
involve very different physics -- nuclear reactions when the Universe
was seconds old, gravity-driven acoustic oscillations when the Universe
was around 400,000 yrs old, and the inhomogeneity in the
distribution of matter in the Universe today -- and thus also provide
an important test the consistency of the big-bang framework
and general relativity.

Next, consider the ratio of the total matter density to the baryon
density:
\begin{eqnarray}
\Omega_M/\Omega_B & = & 7.2\pm 2.1 \qquad {\rm CMB}\nonumber \\
                  & = & 6.7\pm 3.1 \qquad {\rm Power\ Spectrum}\nonumber \\
                  & = & 9.0 \pm 1.4 \qquad {\rm Clusters\ (SZ)}\ +\ H_0\nonumber \\
                  & = & 8.7\pm 1.6 \qquad {\rm Clusters\ (x-ray)}\ +\ H_0\nonumber
\end{eqnarray}
Again, all four measurements are clearly consistent and involve different
physics -- gravity driven acoustic oscillations, inhomogeneity in
the distribution of matter today, and cluster dynamics.

Using standard techniques
(and flat priors) I have constructed the likelihood function
for $\Omega_M$, $\Omega_B$ and $h$.  From this,
a posteriori probability distributions and credible ranges
are calculated by marginalizing over one or two of these
quantities in the usual way.  The $1-\sigma$ ranges are:
\begin{eqnarray}
\Omega_B  =  0.039 \pm & 0.0075& \ \
\Omega_M  =  0.33 \pm 0.035 \nonumber\\
h & = & 0.69 \pm 0.06
\end{eqnarray}

While the one-dimensional probability distributions are not
perfectly Gaussian, the 68\% and 95\% credible ranges
match pretty well with these $1\sigma$ error flags.

The value for the baryon fraction is largely
driven by BBN and $H_0$; alone they
imply $\Omega_B = 0.0385\pm 0.0077$.  Likewise, the Hubble
constant is largely driven by its direct determination.
The different scalings of the baryon-to-matter ratios
with Hubble constant also provide important leverage, which
can be seen in two ways: (1) the joint determination of $h$ has a slightly
smaller uncertainty than the direct measurement alone,
$\pm 0.06$ vs. $\pm 0.07$; (2) if one arbitrarily doubles the
uncertainty in the Hubble constant, $h=0.72\pm 0.14$, and carries
out the same analysis, the results do not change dramatically:
\begin{eqnarray}
\Omega_B  =  0.040 \pm & 0.012 & \ \
\Omega_M  =  0.33 \pm 0.045 \nonumber\\
h & = & 0.69 \pm 0.075
\end{eqnarray}

To investigate the robustness of my estimates for the
matter and baryon densities I have studied
their sensitivity to the individual input data.
To investigate sensitivity to scale-dependent bias on $\Omega_Mh$ I
increased $\sigma_{\Omega_Mh}$ by a factor of 4.  The central
values for $\Omega_M$ and $\Omega_B$ were unchanged and
their errors increased to $\pm 0.044$ and $\pm 0.0076$ respectively.
One by one I doubled the errors for the other input data;
the central values for $\Omega_M$ and $\Omega_B$
changed little and the errors grew to at most $\pm 0.05$ and
$\pm 0.008$ respectively.  One minor trend was observed:
enlarging the error flags on both the X-ray and SZ cluster
baryon fractions by a factor of 4, to significantly reduce
the weight given to the cluster baryon fraction, decreased the estimate
for the matter density by about $1\sigma$, $\Omega_M =0.29\pm 0.05$.

\section{Concluding remarks}

Based upon present measurements of physical quantities not
tied to the distribution of light I conclude
that:  $\Omega_M = 0.33\pm 0.035$ and $\Omega_B = 0.039\pm 0.0075$.
While potential sources of systematic error
remain, I have shown that these mass-density determinations are
robust, and do not depend strongly upon any one measurement.

The precision and reliability of these determinations
should improve over the next decade.
With Planck and MAP the uncertainty in both $\Omega_Mh^2$ and
$\Omega_Bh^2$ is likely to drop to the percent level
(see e.g., Eisenstein et al, 1999);
the cluster and power spectrum derived quantities
are likely to improve by at least a factor of two.  The situation
with the Hubble constant is more difficult to predict.

To illustrate the potential for improvement, I have reduced the
uncertainty in $\Omega_Mh^2$ and $\Omega_Bh^2$ to 1\%, in $\Omega_Mh$
to 5\%, in $\Omega_B/\Omega_M$ to 20\%, and in the cluster baryon
fractions and Hubble constant by a factor of two.  The projected
$1\sigma$ uncertainties for $\Omega_M$ and $\Omega_B$ drop to $\pm 0.012$
and $\pm 0.0017$ respectively.

Already the physically based matter density has interesting
implications.  First, combining it with the CMB determination of
the total mass/energy density, $\Omega_0 = 1 \pm 0.05$ (Hanany et al,
2000; Netterfield et al, 2002; Pryke et al, 2002),
makes a very strong case for an additional
component to the Universe (referred to as dark energy)
that is smoothly distributed with $\Omega_X = 0.67\pm 0.06$.
This evidence for the dark energy bolsters significantly
the direct evidence for dark energy from supernovae
(Perlmutter et al, 1999; Riess et al, 1998).

Second, the physically based matter density is about a
factor of 2 larger than that determined from the mass-to-light ratios
of clusters, $\Omega_M = 0.19\pm 0.04$ (Carlberg et al, 1997)
and $0.16\pm 0.05$ (Bahcall et al, 2000).
This suggests that baryons in the cluster environment
produce twice as much light as in the field.  Going a
step further, this implies the fraction of baryons that become
stars is about a factor of two higher.
The SDSS will determine mass-to-light ratio for a stacked sample
of millions of field galaxies in five-color bands using weak
lensing and will test this hypothesis by providing a reliable estimate
of $\langle M/L \rangle$.

Finally, in probing the nature of the dark energy, independent knowledge
of the matter density is crucial for breaking the degeneracy between
the dark energy equation-of-state ($w_X\equiv p_X/\rho_X$) and $\Omega_M$
(see e.g., Maor et al, 2001; Weller \& Albrecht, 2001; Huterer \&
Turner, 2001).  The value
deduced here is essentially independent of the nature of the dark energy
(assuming a flat Universe), and the current one-sigma uncertainty
would already help significantly to break the degeneracy in future
determinations of $w_X$.

\paragraph{Acknowledgments.}
This work was supported by the DoE (at Chicago and Fermilab), the
NSF (through the CfCP) and by the NASA (at Fermilab by grant NAG 5-7092).
I thank the participants of the 2002 Workshop on Large-scale Structure
at the Aspen Center for Physics for helpful comments.

\end{document}